\documentclass[onecolumn,showpacs,preprintnumbers]{revtex4}%
\usepackage{amsfonts}
\usepackage{amsmath,epsfig}
\usepackage{graphicx}
\usepackage{dcolumn}
\usepackage{bm}
\usepackage{amssymb}
\usepackage{amsmath}%
\begin{document}
\title{Combustion of a hadronic star into a quark star: the turbulent and the diffusive regimes}
\author{Alessandro Drago and Giuseppe Pagliara}
\affiliation{Dip.~di Fisica e Scienze della Terra dell'Universit\`a di Ferrara \\
INFN Sez.~di Ferrara, Via Saragat 1, I-44100 Ferrara, Italy}

\begin{abstract}
We argue that the full conversion of a hadronic star into a quark or a
hybrid star occurs within two different regimes separated by a
critical value of the density of the hadronic phase
$\overline{n_h}$. The first stage, occurring for $n_h
>\overline{n_h}$, is characterized by turbulent combustion and lasts
typically a few ms. During this short time-scale neutrino cooling is
basically inactive and the star heats up thanks to the heat released
in the conversion. In the second stage, occurring for
$n_h<\overline{n_h}$, turbulence is not active anymore, and the
conversion proceeds on a much longer time scale (of the order of tens
of seconds), with a velocity regulated by the diffusion and the
production of strange quarks. At the same time, neutrino cooling is
also active. The interplay between the heating of the star due to the
slow conversion of its outer layers (with densities smaller than
$\overline{n_h}$) and the neutrino cooling of the forming quark star
leads to a quasi-plateau in the neutrino luminosity which, if observed,
would possibly represent a unique signature for the existence of quark
matter inside compact stars. We will discuss the phenomenological
implications of this scenario in particular in connection with the
time structure of long gamma-ray-bursts.
\end{abstract}

\maketitle

\section{Introduction}
The Bodmer-Witten hypothesis on the absolute stability of strange
quark matter \cite{Bodmer:1971we,1984PhRvD..30..272W} stimulated many
interesting investigations on the possible existence of compact stars
entirely composed by this kind of matter
\cite{Alcock:1986hz,Haensel:1986qb} or small nuggets of strange quark
matter which would propagate in the Universe as cosmic rays
\cite{Madsen:1989pg}. The exothermic process of conversion of ordinary
nuclear matter into strange quark matter has been studied for the
first time in Ref. \cite{1987PhLB..192...71O} where it has been
modeled as a slow combustion by means of a one dimensional stationary
reaction-diffusion-advection equation for the strange quarks
concentration. This is a kinetic theory calculation in which the
microphysical processes occurring within a finite width combustion
zone are taken into account and which allows to determine the velocity
of the conversion as a function of the quarks diffusion coefficient
and the rate of conversion of down quarks into strange quarks. It
turns out that typical values for the burning velocities are within
$10^3-10^4$ cm/sec for a quark chemical potential $\mu_q \sim 300$ MeV
and a temperature of the quark phase $T \sim 10$ MeV.

The limit of this kinetic theory approach is that it does not allow to
take into account possible macroscopic collective flows and
hydrodynamical instabilities driven by pressure and density gradients
between the fuel and the ashes fluids. Gravity would also play an
important role in the dynamics of the conversion.  For a complete
treatment of the problem one would have to couple the equations of
hydrodynamics (i.e. the equations of conservation of baryon number,
momentum and energy) and the equation of conservation of chemical
species (which includes the diffusion and the reaction rates within
the combustion zone) in multidimensional numerical simulations, see
\cite{williams,niebergal2010a}.  The width of the combustion zone
$\delta$ can be estimated by the simple relation $\delta \sim
\sqrt{D\tau}$ \cite{landau} where $D$ is the quark diffusion
coefficient ($D \sim 10^{-1}$cm$^2$/sec for $\mu_q \sim 300$ MeV and
$T \sim 10$ MeV \cite{niebergal2010a}) and $\tau$ is the inverse of
the rate of conversion of down quarks into strange quarks ($\tau \sim
10^{-9}$ sec for $\mu_q \sim 300$ MeV \cite{Alford:2014jha}).  One
obtains $\delta \sim 10^{-5}$ cm. Clearly, it would be numerically
unfeasible to resolve such a small length scale within a numerical
simulation aiming at studying a compact star whose size is of the
order of ten km.

A similar problem exists in the context of numerical simulations of
type Ia Supernovae where length scales from $10^{-4}$ to $10^{8}$ cm
characterize the physical system \cite{Reinecke:1998mk} and an
alternative scheme has been devised: being the combustion zone much
smaller than the size of the system one can assume that it is actually
an infinitely thin layer and it can be treated as a surface of
discontinuity which separates the ashes from the fuel. In this scheme,
called {\it flamelet regime} \footnote{The flamelet regime is also
  characterized by the fact that the large turbulent eddies do not
  alter the structure of the flame at the microscopic level.}, one has
to impose the Hugoniot jump conditions to relate the thermodynamical
variables of the fluid at both sides of the discontinuity. This
approach, implemented in 3+1D, is one the most used in the context of
type Ia Supernovae and it has demonstrated the crucial role played in
such explosive events by the hydrodynamical instabilities,
specifically the Rayleight-Taylor instability and the Landau-Darrieus
instabilities \cite{Hillebrandt:2000ga,blinnikov}, which turns the
laminar combustion into a much faster turbulent combustion. In this
sense the flamelet approximation is very sensible: since the burning
velocity is strongly enhanced by turbulence, the importance to know
the exact value of the laminar velocity (governed by the microphysics
of the combustion zone) is subordinate.

The same method has been adopted for studying the conversion of
nuclear matter into strange quark matter in semi-analytical
calculations
\cite{Lugones:1994xg,Horvath:2007tv,Horvath:2010sc,Drago:2005yj}.
This process is most likely a deflagration (subsonic combustion) and
not a detonation (supersonic combustion) although its velocity is
substantially increased, up to two orders of magnitude, by the
development of hydrodynamical instabilities. A common finding within
this approach is that at some critical density this rapid combustion
stops and the neutron star cannot fully convert into a strange quark
star. These results have been recently confirmed in 3+1D hydrodynamics
numerical simulations: the star is converted on a time scale of ms
thanks to hydrodynamical instabilities but a sizable fraction of the
star, of the order of few $0.1 M_{\odot}$, is left unburnt
\cite{Herzog:2011sn,Pagliara:2013tza}. A natural question that we will
address in this paper arises: does the combustion really stop or does
it proceed on a longer time scale?

After clarifying the reason for which, in a purely hydrodynamical
approach, the turbulent combustion stops (the critical density being
given by the condition proposed by Coll in Ref.\cite{coll}), we will
model the subsequent slow conversion process (which is driven by
strangeness diffusion and weak interaction processes). We will
demonstrate that in such a two steps scenario, a first stage of fast
turbulent combustion and a second stage of slow laminar combustion,
one could obtain a noticeable imprint on the neutrino signal emitted from
the formation of a strange quark star.

The paper is organized as follows: in Section II we will review the
procedure adopted for treating the combustion within a purely
hydrodynamical approach and we will explain the meaning of the Coll's
condition. In Section III we will model the slow combustion process
and provide some example of the neutrino signals released by a strange
star at birth. We draw our conclusion in Section IV.


\section{Turbulent hydrodynamical combustion}
We review here the way the combustion is treated if the combustion
zone is so thin to be considered as a surface of discontinuity,
usually called {\it flame front}. The generalization of the classical
combustion theory within relativistic (magneto-)hydrodynamics has been
presented by Coll and Anile in Refs.\cite{coll,anile}. Denoting with $p_i$,
$e_i$, $n_i$ and $X_i=(e_i+p_i)/n_i^2$ the pressure, energy density,
baryon density and dynamical volume of fluid $i$, the so-called
``condition for exothermic combustion'', we will name it ``Coll's
condition'', for the conversion of fluid $1$ into fluid $2$ reads:
$e_1(p,X)>e_2(p,X)$, i.e. at fixed pressure and dynamical volume, the
energy density of fluid 1, the fuel, must be larger then the one of
fluid 2, the ash, see also \cite{zeldo} for the case of classical
hydrodynamics.
Introducing the enthalpy density $w_i=e_i+p_i$, the Coll's condition
can be equivalently expressed as $w_1(p,X)>w_2(p,X)$. {\it As we will
  show, this condition is necessary in the case of a detonation, while
  in the case of a deflagration it determines the appearance of
  hydrodynamical instabilities dramatically accelerating the
  combustion process.} To our knowledge this issue is not discussed in
any standard textbook of hydrodynamics nor in any research paper. Our
result also implies that Coll's condition corresponds to the request
of exothermicity in the case of detonations while its violation does
not exclude the possibility of slow processes of combustion.

\begin{figure}[ptb]
\vskip 1cm
\begin{centering}
\epsfig{file=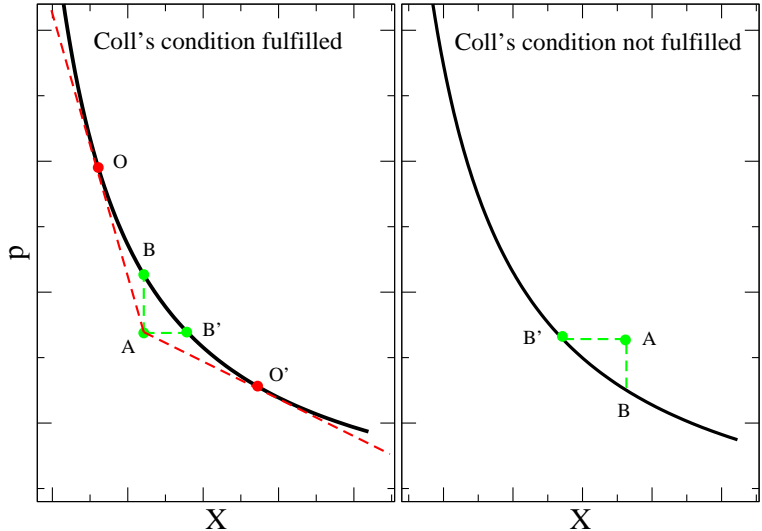,height=10cm,width=7cm,angle=-90}
\caption{Illustrative plot of the detonation adiabat in the case in which the Coll's 
condition is fulfilled (left panel) or not (right panel).
A, B, B' indicate  respectively the initial hadronic state and two possible final states for the quark phase. O and O'
are the Chapman-Jouget points.}
\label{uno}
\end{centering}
\end{figure}

Let us consider the two fluids to be hadronic matter and quark matter.
For the hadronic matter we adopt a generic equation of state at zero
temperature $e_h(p_h,X_h)$ (this situation corresponds to the case of
the conversion of a cold hadronic star) while for quark matter we
consider the simple case of the equation of state of the MIT bag model
with massless quarks as in Ref. \cite{Herzog:2011sn}. The relation
between energy density and pressure in this case reads: $e_q=3p_q+4B$
where $B$ is the usual bag constant (notice that in the case of
massless quarks the energy density is a function of only $p$ and not
of $X$). The more general case of a polytrope is discussed in the
appendix A, see also \cite{Drago:2015gsa}.

Similarly to the case of the discontinuity associated with a shock
wave, also in the case of the flame front, one has to impose the
continuity equations for the fluxes of baryon number (or mass flux),
momentum and energy. By indicating with $j$ the number of baryons
ignited per unit time and unit area of the flame front,
one obtains the following relations between the thermodynamical
quantities of the hadronic fluid and of the
quark fluid:
\begin{eqnarray}
n_hu_h&=&n_qu_q=j\\
(p_q-p_h)/(X_h-X_q)&=&j^2\\
w_h(p_h,X_h)X_h-w_q(p_q,X_q)X_q&=&(p_h-p_q)(X_h+X_q)
\end{eqnarray}
the last equation corresponding to the so-called relativistic
detonation adiabat. $u_h$ and $u_q$ are the four-velocities of
hadronic and quark matter in the flame front rest frame.
Starting from hadronic matter in the state A: $p_h=p_A$ and $X_h=X_A$
and for a given value of $j$, Eqs. 1-3 allow to determine the final
state B of quark matter, $p_q=p_B$ and $X_q=X_B$ which lies on the
detonation adiabat. In particular, the second equation represents a
line in the (p,X) plane passing through A and with angular coefficient
equal to $-j^2$.  The intersections of this line with the detonation
adiabat allow to find the state B of quark matter. The baryon number
flux $j$, or equivalently the flame front velocity with respect to one
of the two fluids $u_i$, in general cannot be expressed in terms of
the thermodynamical variables of the states A and B. It is instead
related to the specific microscopic properties of the chemical
reactions of the combustion, the heat transfer and the diffusion of
chemical species across the flame front
and therefore it must be determined within a kinetic theory approach
such as the one of Ref. \cite{1987PhLB..192...71O}. We remark however
that hydrodynamical instabilities can substantially affect the flame
front velocities as we will discuss later on.

We will now show that Coll's condition determines the position 
of the initial state A with respect to the detonation adiabat.
To this purpose we will consider a point on the detonation adiabat having 
the same value of X of the initial state A ($X_A=X_B$) and we will compute
the corresponding value of the pressure $p_B$ (this procedure corresponds to the limit 
$j \rightarrow \infty$): in this way we will find if A lies below or above the detonation adiabat.
Coll's condition, written in terms of the enthalpy density, 
reads \cite{coll}: $\Delta(p,X)=e_h(p,X)-e_q(p,X)=w_h(p,X)-w_q(p,X)>0$.
One can write Eq.3, adding and subtracting $e_q(p_A,X_A)$, as:
\begin{eqnarray}
w_h(p_A,X_A)-w_q(p_B,X_A)&=&2p_A-2p_B\nonumber \\e_h(p_A,X_A)+p_A-e_q(p_B,X_A)-p_B+e_q(p_A,X_A)-e_q(p_A,X_A)&=&2p_A-2p_B \nonumber \\ 
\Delta(p_A,X_A)+e_q(p_A,X_A)-e_q(p_B,X_A)&=&p_A-p_B\nonumber \\
\Delta(p_A,X_A)+3p_A+4B-3p_B-4B &=& p_A-p_B \nonumber \\ \Delta(p_A,X_A)=2(p_B-p_A)
\end{eqnarray}
Therefore if $\Delta(p_A,X_A)>0$, i.e. if the Coll's condition is
fulfilled, then $p_B>p_A$ which means that the initial state of the
hadronic phase A lies in the region of the (p,X) plane below the
detonation adiabat. In turn, this implies that there are two specific
values of $j$, $j_O$ and $j_{O'}$, for which the lines passing through
A are tangent to the detonation adiabat. The two points of tangency
are the Chapman-Jouget points, see Fig.1. Following the standard
treatment, one could obtain strong and weak detonations and strong and
weak deflagrations depending on the specific values of $j$ and on the
boundary conditions of the problem. The point O corresponds to the
Chapman-Jouget detonation and it is the only possible realization of
detonation in a physical system, such a compact star, in which no
external force is producing the shock wave, see\cite{landau}.
Moreover a detonation taking place in a compact star can be
assimilated to a detonation in a closed pipe and also in this case it
can take place only at the Chapman-Jouget point \cite{landau}.
Clearly, if the Coll's condition is not fulfilled, the state A lies in
the region of the (p,X) plane above the detonation adiabat (see Fig.1
right panel) and there are no Chapman-Jouget points in this
case. Detonation is therefore excluded when the Coll's condition is
not fulfilled.

Let us discuss now how the Coll's condition is related to the
deflagration regimes. We consider the simplest case of a slow
combustion, i.e. a process in which the velocities $v_h$ and $v_q$ are
much smaller than the sound velocities $c_h$ and $c_q$ of the two
fluids (this case is particularly interesting for the combustion of
hadronic stars into quark stars for which the laminar velocities,
found in \cite{1987PhLB..192...71O}, are much smaller than the sound
velocities). By using Eqs.(1-3) one finds in this regime that
$p_h=p_A=p_q=p_{B'}$ and $w_A/n_A=w_{B'}/n_{B'}$ or equivalently
$(e_A+p_A)/n_A=(e_{B'}+p_A)/n_{B'}$, i.e. the enthalpy per baryon is
conserved during the combustion (see \cite{landau} for the case of
non-relativistic hydrodynamics).  One sees immediately that in this
case the Coll's condition implies, see Fig.1 (left panel), that
$X_{B}'>X_A$ i.e.  $(e_{B'}+p_A)/n_{B'}^2 > (e_A+p_A)/n_A^2$ which
together with the conservation of the enthalpy per baryon implies
$n_{B'}<n_A$. Moreover, from
$n_A(e_{B'}+p_A)=n_{B'}(e_A+p_A)<n_A(e_A+p_A)$ on obtains
$e_{B'}<e_A$.  Thus the quark phase is produced with baryon density
and energy density smaller than the one of the hadronic phase. This
fact opens the possibility of obtaining the Rayleigh-Taylor
instabilities.  These instabilities arise, in the presence of gravity,
if the gradient of the gravitational potential and the gradient of the
energy density point in opposite directions (“inverse density
stratification”). Indeed as shown in
Refs. \cite{Drago:2005yj,Herzog:2011sn}, the Rayleight-Taylor
instabilities do occur during the conversion of an hadronic star and
they substantially increase the efficiency of burning leading to time
scales of the order of ms for the conversion of a big portion of the
star.

On the other hand, if the Coll's condition is violated, the new phase
is produced (again in the case of a slow combustion) with $e_{B'}>e_A$
and therefore the burning can proceed but with velocities which are
dominated by the diffusion and the rate of the chemical reactions and
which are therefore much smaller than the velocities obtained during
the turbulent regime.  Moreover one can notice that since the new
phase is produced with an energy density larger than the one of the
fuel one cannot speak in this case of a deflagration (for which the
inequality $e_{B'}<e_A$ must hold true).

One can separate the turbulent from the diffusive regime by finding
the critical density of the hadronic phase, $\overline{n_h}$, for
which the Coll's equality is satisfied $e_h(p,X)=e_q(p,X)$. For this
value of density the state A lies on the detonation adiabat. Moreover one also
obtains that $n_q=\overline{n_h}$: thus pressure, energy density and
baryon density are continuous across the interface of the flame front.
The evolution of the flame for $n_h<\overline{n_h}$ is not turbulent
anymore.

A simple way to understand the role played by the instabilities is to
use the analytic model proposed in \cite{blinnikov} which allows to
estimate the increase in the laminar velocities $u_i$ due to
turbulence in terms of the fractal dimension of the flame front. In
particular, the fractal excess of the wrinkled front surface $\Delta
D$ is proportional to the square of $\Gamma=1-e_B/e_A$ and it
therefore vanishes when the Coll's condition is met at
$n_A=\overline{n_h}$. The mean velocity $v_{mh}$ of the front (with
respect to the hadronic fluid) is significantly larger than the
laminar velocity $v_{lh}$ and it reads:
\begin{equation} 
v_{mh}=v_{lh} (\lambda_{\mathrm{max}}/\lambda_{\mathrm{min}})^{\Delta D}
\label{wrinkle}
\end{equation}

where $\lambda_{\mathrm{max}}$ and $\lambda_{\mathrm{min}}$ are two
length scales which regulate the maximal and the minimal size of the
wrinkle for which the velocity of the Rayleight-Taylor growing modes
is larger than the laminar velocity, see \cite{Drago:2005yj}.  As one
can indeed notice in the simulations of \cite{Herzog:2011sn}, the
velocity of the flame front approaches the laminar velocity when the
density is close to $\overline{n_h}$. Clearly a fraction of the star
cannot be converted by means of a fast hydrodynamical deflagration and
actually a mass of a few $0.1 M_{\odot}$ remains unburnt at the end of
the turbulent regime \cite{Herzog:2011sn}.

\begin{figure}[ptb]
\vskip 1cm
\begin{centering}
\epsfig{file=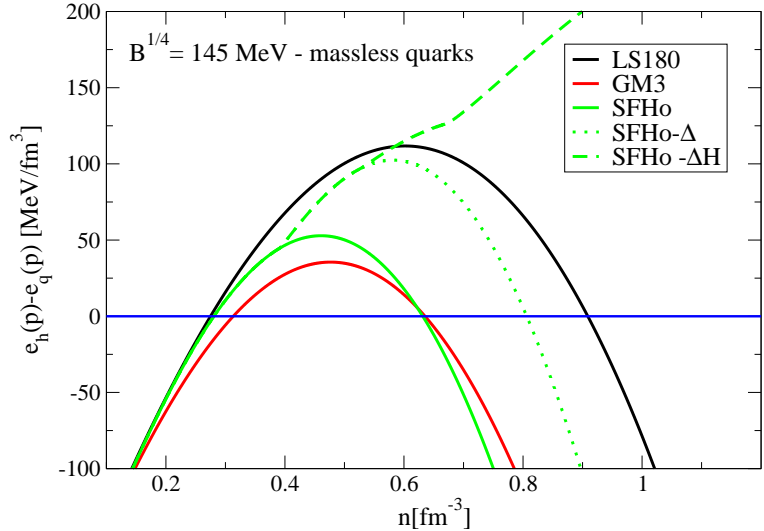,height=10cm,width=7cm,angle=-90}
\caption{Difference between the energy density of the hadronic phase and the quark phase as a 
function of the baryon density. The solid black and red lines correspond to the LS180 and GM3 equations of state
with only nucleons. The green lines correspond to the SFHo model with only nucleons (solid line),
deltas (dotted line) and deltas and hyperons (dashed line). 
The Coll's condition is fulfilled
within the density window for which this difference is positive.}
\label{due}
\end{centering}
\end{figure}

To provide some example of the results obtained by imposing the Coll's
condition, we display in Fig.2 the difference between the energy
density of hadronic matter and of quark matter (again massless quarks for
simplicity) for the following equations of state: the LS180 model from
\cite{Lattimer:1991nc}, the GM3 model from \cite{Glendenning:1991es}
and the SFHo model from \cite{Steiner:2012rk} also with the inclusion
of delta resonances and hyperons \cite{Drago:2014oja}. Pure nucleonic
equations of state provides two values of baryon density for which the
Coll's equation holds (as also found in
\cite{Lugones:1994xg,Herzog:2011sn}). In general, the stiffer the
hadronic equation of state the smaller the density window for which
the turbulent hydrodynamical combustion can take place. One the other
hand, when considering deltas and hyperons, within the SFHo model, the
turbulent hydrodynamical combustion is always possible above the
threshold for the formation of deltas (see dashed line). Notice that
once a certain amount of hyperons is present in a hadronic star, the
conversion is a necessary process in the scenario of the Witten's
hypothesis \cite{Drago:2013fsa}.

\begin{figure}[ptb]
\vskip 1cm
\begin{centering}
\epsfig{file=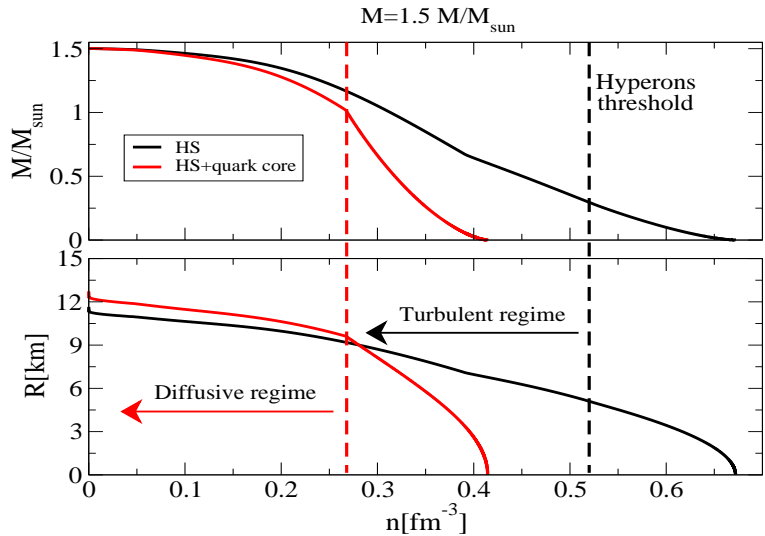,height=10cm,width=7cm,angle=-90}
\caption{Enclosed gravitational mass and radius as a function of the baryon density for a $1.5 M_{\odot}$ 
hadronic star before the turbulent conversion (black lines) and after the turbulent conversion (red lines). The black dashed line marks the appearance 
of hyperons: the seed of strange quark matter is formed
at densities larger than this threshold. The red dashed line marks the density 
below which Coll's condition is no more fulfilled
and the turbulent combustion does not occur anymore. Below this density, the combustion proceeds
via the slow diffusive regime.}
\label{tre}
\end{centering}
\end{figure}

\section{Diffusive regime}
\subsection{Diffusion of strangeness and exothermicity}
As explained before, when the flame front reaches $n_h=\overline{n_h}$
the hydrodynamical instabilities responsible for the very fast
conversion of the inner part of the star are not active anymore. The
subsequent evolution of the system can be described by using the
laminar burning velocity of the front $v_{lh}$ computed in
\cite{1987PhLB..192...71O,Alford:2014jha}: it corresponds to $v_{mh}$
in the limit $\Delta D \rightarrow 0$ (see Eq. \ref{wrinkle}).  Let us
first review how the conversion process is described within
microscopic kinetic theory approaches: we follow in particular the
more recent treatment presented in \cite{Alford:2014jha}. The
conversion is basically due to two simultaneous processes: diffusion
of quarks within the combustion layer and flavor changing weak
interactions among quarks (the process $u+d \rightarrow u+s$ being the
most relevant for our case). One starts by introducing a position
dependent strangeness unbalance $a(x)=(n_K(x)-n_K^Q)/n_Q$
where the
coordinate $x=0$ defines the position of the conversion boundary,
$n_K(x)=(n_d(x)-n_s(x))/2$ and $n_K^Q=n_K(x)$ for $x
\rightarrow +\infty$ (it would vanish for instance in the case of
massless strange quarks). In the limit $x \rightarrow +\infty$ one has
bulk equilibrated strange quark matter with baryon density $n_Q$,
while for $x<0$, $a(x)=\mathrm{cost}=a(0^-)\equiv a_N$. If the hadronic
phase is assumed to be made only of neutrons and if $n_K^Q=0$ then
$a_N=n_N/n_Q$ with $n_N$ being the baryon density of neutron matter.
In the following, for simplicity, we will adopt these conditions.
Within the combustion zone, where both
diffusion and chemical reactions are active, strange quark matter is
out of beta equilibrium. There are two useful reference frames: the one in which 
the flame front is at rest and the one in which quark matter is at rest.
If $v_{lh}$ and $v_{lq}$ are the (laminar) velocities of hadronic matter and
quark matter in the front frame, in the quark matter
frame the velocity of the front is $v_{lf}=-v_{lq}$ and
$v_{lh}'=v_{lh}-v_{lq}$.
We remind that $v_{lq}$ and $v_{lh}$
are related by the baryon flux continuity equation: $v_{lq} n_q=v_{lh} n_h$.

By solving the steady state
transport equation for $a$ given in \cite{1987PhLB..192...71O,Alford:2014jha}
with the boundary conditions $a(0^+)\equiv a_{Q^*}=n_K^Q(0^+)/n_Q$ and $a(x\rightarrow +\infty)=0$ (notice that $a(x)$ monotonically decreases
from $x=0$ to $x \rightarrow +\infty$)
one obtains for the velocity of the front in the quark matter rest frame (in the suprathermal regime of Ref.\cite{Alford:2014jha}):

\begin{equation}
v_{lf}=\sqrt{\frac{D}{\tau}\frac{a_{Q*}^4}{2a_N (a_N-a_{Q*})}}.
\end{equation}

The quark diffusion coefficient $D$
and the inverse rate of chemical reactions (for the process $u+d \rightarrow u+s$) $\tau$ are given by \cite{niebergal2010a,Alford:2014jha}:
\begin{equation}
D=0.1\left(\frac{\mu_q}{300\, \mathrm{MeV}}\right)^{2/3}\left(\frac{T}{10\, \mathrm{MeV}}\right)^{-5/3}\mathrm{cm^2/sec},\,\,\,\,\,\, 
\tau=1.3 \times 10^{-9} \left(\frac{300\, \mathrm{MeV}}{\mu_q}\right)^5\mathrm{sec}.
\end{equation}
Notice that $v_{lf}$ scales as $\sim T^{-5/6}$: the more the star is heated up 
by the conversion process the slower the burning velocity. This fact will 
have important consequences for the thermal evolution of the star
and the neutrino signal which will be discussed in the following.
The value of the parameter $a_{Q*}$ is crucial: 
the larger $a_{Q*}$ the larger $v_{lf}$.
This quantity is just the boundary condition for the transport equation
but its value cannot be taken arbitrarily large because one needs
to impose the exothermicity of the process of combustion.
In other terms, if the unbalance $a_{Q*}$ is too large
the front cannot move because there are not enough strange quarks
to trigger deconfinement. Its maximum value $a_{Q*}^{max}$
is estimated in Appendix B.

Let us now explain how we model the conversion during the diffusive
regime.  Once the initial state A of the hadronic phase is fixed, at a
density $n_h \leq \overline{n_h}$, one needs two equations to
determine the state of the newly produced quark phase (for instance in
terms of $\mu_q$ and $T$). As in the case of a weak deflagration discussed before, 
since the velocities during the diffusive regime are
small with respect to the sound velocities, we can set
$p_h=p_A=p_q=p_{B'}$ (as also assumed in
\cite{1987PhLB..192...71O,Alford:2014jha}) and moreover
$w_h/n_h=w_A/n_A=w_q/n_q=w_{B'}/n_{B'}$ \footnote{One obtains this
  relation by using again the continuity equations at the front
  interface under the hypothesis of small velocities. Alternatively, it
  can be obtained from the first principle of thermodynamics applied
  to a system in which a number $\Delta N_1$ of baryons of fluid 1 are
  converted to $\Delta N_2$ baryons of fluid 2 at a constant pressure
  $p$: the first principle of thermodynamics reads in this case
  $\Delta Q= p \Delta V_1+ \Delta U_1+p \Delta V_2+\Delta U_2$.  One
  can write $\Delta V_i=\Delta N_i/n_i$ and $\Delta U_i=e_i/n_i \Delta
  N_i$.  From the conservation of baryon number $-\Delta N_1=\Delta
  N_2>0$. Moreover, in a purely hydrodynamical framework no heat is
  assumed to be transported out of the system and therefore $\Delta
  Q=0$.  This leads to $w_1/n_1=w_2/n_2$ i.e. to the conservation of
  the enthalpy per baryon.}
Notice that these two equations match continuously at $n_h =\overline{n_h}$ with the equations for the turbulent hydrodynamical
regime. 

The process of conversion is exothermic if during the temporal
evolution of the system some heat is released to the environment.  Let
us discuss how the heat per baryon generated by the conversion can be
calculated. The equation of conservation of the enthalpy per baryon
$w_A/n_A(p_A,T_A)=w_B/n_B(p_A,T_B)$ allows to find the value of $T_B$
and if it turns out that $T_B>T_A$, the process is exothermic. 
By indicating with $N$ the number of
baryons composing the system, the total initial enthalpy is given (for
uniform matter) by $N w_A/n_A(p_A,T_A)$.  Because of cooling,
asymptotically the system will reach again the initial temperature $T_A$
and the total enthalpy of the system after the full
conversion and after the cooling process will be $N
w_B/n_B(p_A,T_A)$. Hence, the total heat $Q$ released by the process
of conversion and emitted into the environment (via neutrino cooling
in the case of a compact star) is given by $Q\equiv
N(w_B/n_B(p_A,T_B)-w_B/n_B(p_A,T_A))=
N(w_A/n_A(p_A,T_A)-w_B/n_B(p_A,T_A))$. Thus, the difference between the
enthalpy per baryon of the fuel and of the ashes calculated at the same pressure and temperature
corresponds to the heat per
baryon $q$ released by the conversion:
$q=w_A/n_A(p_A,T_A)-w_B/n_B(p_A,T_A)$.

A natural question concerns the point at which the conversion will
stop. Let us indicate with $p_h(r)$ and $p_q(r)$ the hadronic
matter and quark matter pressure profiles inside the star ($r$ is the
radial coordinate). The radius of the star $R$, is obtained by
imposing $p(r)=0$ whatever is the composition of the surface of the
star.  When the burning front is close to the surface of the star, the
conservation of the enthalpy per baryon (for an initially cold star)
implies
$e_A/n_A(T_A=0,p_A=0)=e_B/n_B(T_B>0,p_A=0)>e_B/n_B(T_B=0,p_A=0)$.  But
$e_A/n_A(T_A=0,p_A=0)=930$ MeV (corresponding the energy per baryon of
the iron nuclei composing the outer crust) and it is necessarily
larger than $e_B/n_B(T_B=0,p_A=0)$ under the hypothesis of absolutely
stable strange quark matter. This means that the conversion is
exothermic, and therefore it will continue, up to the surface of the
star. Therefore, just by thermodynamics arguments, one would expect
that the whole hadronic star converts into a quark star.  On the other
hand, the existence of a crystalline structure in the outer crust can
significantly hinder the conversion process
\cite{1987PhLB..192...71O}.  It has been noticed in
Ref.~\cite{Haensel:2007tf} that the preheating of the crust due to the
heat released by the conversion of the inner region leads to the
dissociation of nuclei which greatly facilitates the burning into
quark matter. In \cite{Haensel:2007tf} it has been estimated that the
time needed to dissolve and convert the outer crust is of the order of
few tens of ms. We will neglect
in our calculation the
conversion of the outer crust and consider only the, slower,
conversion of the hadronic layer between $\overline n_h$ and the
neutron drip line density $n_d=2.6\times 10^{-4}$fm$^{-3}$.
The conversion of the outer layer can be important e.g.
when discussing short gamma-ray-bursts. We will consider this problem
in a future paper.

\subsection{Solving the diffusion equations}
We want now to discuss an example of the temporal evolution of the
burning of a hadronic star during the diffusive regime.  As a first
ingredient, we need an initial profile for the star after the first
stage of turbulent burning: this configuration is composed by hot
quark matter for densities larger than $\overline n_h$ and by cold
hadronic matter for densities smaller than $\overline n_h$. The
equation of state of hot quark matter is computed as explained above
by requiring that at fixed pressure, the enthalpy per baryon of the
quark phase is equal to the one of the hadronic phase as in the case
of a slow combustion.  This assumption is not correct for the
turbulent regime, during which the conversion velocity is
significantly increased by the hydrodynamical instabilities; one can
expect that the kinetic energy of the fluid flow is completely
dissipated into heat once turbulence is over.

We start from $1.5 M_{\odot}$ hadronic star obtained with the SFHo
model with the inclusion of deltas and hyperons.  The central density
of this stellar object is larger than the critical density for the
formation of hyperons. In Fig.3, the black lines correspond to the
enclosed gravitational mass (upper panel) and the radius (lower panel)
as a function of the baryon density for the initial hadronic star
configuration. The black dashed line indicates the onset of hyperons
which are needed, in our scenario, to seed quark matter.  The red
lines correspond to the hybrid configuration, hot quark core and cold
hadronic layer, after the turbulent regime. For the quark matter
equation of state we have adopted the MIT bag model with massive
strange quarks and with the inclusion of the perturbative QCD
corrections (parameter set 1 of Ref.\cite{Pagliara:2013tza}). Finally,
the red dashed line located at $n=\overline n_h$ separates the
turbulent from the diffusive regime. During the conversion the total
baryonic mass, which turns out to be $M_B=1.71 M_{\odot}$, is
conserved.  Notice that the total gravitational mass is also conserved
during the turbulent conversion: this is due to the fact that cooling
is not active during this stage and can be neglected. After the
turbulent regime, a layer of hadronic matter of about $0.5M_{\odot}$
with a width of about $3$km is left unburnt. This situation is exactly
the one obtained also in the numerical simulation of
Ref.\cite{Pagliara:2013tza} and depicted in Figs 2 and 3 of that
paper.

Let us now introduce the two differential equations
describing the propagation of the flame front and the thermal evolution 
of the star. We work in the reference frame of quark matter. 
Concerning the position of the flame front, by labeling 
with $r_f(t)$ its radial coordinate, one can write:
\begin{equation}
\frac{\mathrm{d}r_f}{\mathrm{d}t}=v_{lf}(\mu_q,T)
\label{diff2}
\end{equation}
with the initial condition $r_f(0)=\overline{r}$ where $\overline{r}$ is obtained
from the baryon density profile by using the
equation $n_h(\overline{r})=\overline{n_h}$.

For handling the thermal evolution of the star, in principle, one
should couple Eq.\ref{diff2} with a partial differential equation
describing the heat transport through neutrinos and the heat source
term related to the energy released by the conversion. Moreover one
should also introduce the thermal conductivity which determines how
the new release of energy is distributed within the star. The effect
of gravity should also be considered to take into account the
readjustment of the star during this stage.  This is a very
complicated task that we will not face in this work.  We will adopt
instead some simplifying assumptions that allow us to understand
qualitatively how the conversion process proceeds and to obtain some
order of magnitude estimates.

First we can notice that the equations describing heat diffusion (and
also the readjustment of the star) have been numerically solved in
Ref.\cite{Pagliara:2013tza}.  In that paper the diffusion regime was
not discussed, but since strangeness diffusion is a rather slow process
we can assume that the first few seconds of the thermal evolution of
the star are dominated by the diffusion of the heat deposited during
the rapid burning of its central region.  We therefore assume that the
temperature of the surface of the star and the neutrino luminosity
evaluated in \cite{Pagliara:2013tza} are the correct ones during the
first seconds, even if the slow conversion of the outer layer is not
taken into account.  From Fig. 3 of Ref.\cite{Pagliara:2013tza} one
can also notice that after a few seconds the thermal profile inside
the star flattens and that in particular the outer region
($r>\overline{r}$) is almost isothermal after about 7-8 seconds.  This
first contribution to the neutrino luminosity can be approximated by
using the simple formula
\begin{equation}
L(t)=Q/\tau e^{-t/\tau}
\label{formulina}
\end{equation} 
where $Q$ is the total 
energy released during the rapid conversion and $\tau$ is the time-scale
of neutrino diffusion \cite{shapiro}. The results of  \cite{Pagliara:2013tza} are well
approximated by taking $\tau \sim 3$ s and $Q \sim 8.5\times 10^{52}$ erg.

\begin{figure}[htb]
\vskip 2cm
\begin{centering}
\epsfig{file=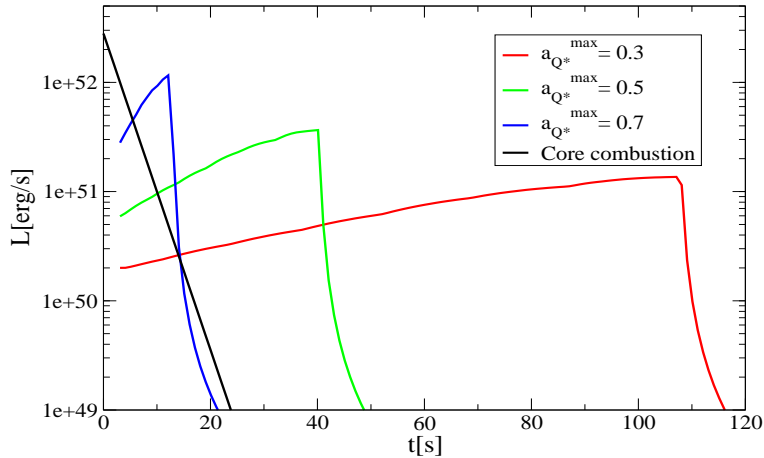,height=10cm,width=6cm,angle=-90}
\caption{Neutrino luminosity associated with the burning during the diffusive regime of the combustion 
for three choices of the parameter $a_{Q*}^{max}$. The black line represents the luminosity obtained from the 
rapid combustion of the core as estimated by using Eq.\ref{formulina}.}
\label{luminosity}
\end{centering}
\end{figure}

To simplify the description of the conversion of the outer layer we
assume that: i) the conduction of heat is extremely fast (infinite
thermal conductivity) in comparison with the burning velocity and
therefore the heat generated by the conversion is distributed
throughout the star (the very same assumption has been made in
\cite{1987PhLB..192...71O}); ii) we assume that neutrinos are
completely trapped and only the black body surface emission is
considered for the cooling. The corresponding luminosity reads $L=21/8
\sigma (T/K)^4 4 \pi r_s^2$ erg/s \cite{shapiro} with $r_s$ the
radius of the neutrinosphere (we will assume that it is located at the
interface between the inner crust and the outer crust where
$n_h=n_d$.)
These two assumptions imply that the temperature is uniform within the
star and one can write a simple equation that expresses the energy
conservation:

\begin{equation}
C(T)\frac{\mathrm{d}T}{\mathrm{d}t}=-L(T) + 4\pi r_f^2\, j(r_f,T)\, q(r_f,T)
\label{diff-full}
\end{equation}
where $C$ is the heat capacity of the star, $L$ the neutrino
luminosity and $j=n_h v_{lh}'$ is the number of baryons ignited per
unit time and unit area. The thermodynamical variables $n_h$, $q$ and
$\mu_q$ (which appears in $v_{lh}'$) are all functions of $r_f(t)$ and
$T(t)$. Concerning the heat capacity, we use $C=2\times
10^{39}M/M_{\odot}(T/10^9)$ erg/K obtained in Ref. \cite{shapiro} for
a uniform density quark star or a hadronic star.  

By solving simultaneously Eqs. \ref{diff2} and \ref{diff-full} with
initial conditions: $r(0)=\overline{r}$, $T(0)=T_0$ MeV (which is the
temperature of the star for $r>\overline{r}$ after the turbulent
regime and it is of the order of $5$ MeV as found in
\cite{Pagliara:2013tza})
we can calculate the time needed to complete the conversion of the
star and the neutrino luminosity due to the conversion of the material
left unburnt after the turbulent stage. In Fig.\ref{luminosity}, we
show three cases corresponding to different values of $a_{Q*}^{max}$ ($a_N$
is fixed to one as in \cite{1987PhLB..192...71O}).  We also display
the curve of luminosity corresponding to Eq. \ref{formulina}.

As discussed above, we assume that the neutrino luminosity estimated
in \cite{Pagliara:2013tza} and approximated by Eq.  \ref{formulina} is
close to the exact one during the first seconds.
After some 10 seconds the luminosity estimated by taking into account
the conversion of the outer layer becomes larger than the one obtained
in \cite{Pagliara:2013tza} where the outer layer remained
unburnt. From that moment forward we assume that the estimate for the
neutrino luminosity obtained by solving Eqs. \ref{diff2} and
\ref{diff-full} is close to the exact one. In other words, we have
separated the complicated problem of the heat diffusion and of the
burning of the outer layer into two stages: the first one lasting a
few seconds during which the neutrino luminosity is dominated by the
heat deposited during the rapid combustion of the core and is
evaluated by solving the problem of heat transport as in
\cite{Pagliara:2013tza} and a second stage, starting after a few
seconds, during which the process of exothermic conversion of the
outer layer provides enough heat to dominate the temperature of the
star and the neutrino luminosity. During this second slow stage we
have assumed the star to be isothermal.

A quasi-plateau in the neutrino luminosity associated with the
combustion of the hadronic layer is obtained (particularly evident for
the smallest value of $a_{Q*}^{max}$). This feature is a necessary
consequence of the temperature dependence of the burning velocity: as
the conversion proceeds, the temperature increases due to the release
of energy and therefore the velocity decreases. It is a
self-regulating mechanism which rapidly leads to an almost constant
velocity of burning and an almost constant luminosity of
neutrinos. The process goes on until the whole star is converted. The
kink appearing in the luminosity curves signals the end of the
conversion: the following evolution is governed only by the cooling
and the standard power law luminosity is obtained.


\section{Conclusions}
The conversion of a hadronic star into a strange quark star 
and the related neutrino emission can be divided into three
different stages:

\begin{itemize}
\item turbulent conversion of the inner part of the star, $r<\overline{r}$, lasting a few ms. During this phase
cooling is negligible due to neutrino trapping.

\item Diffusive conversion which burns the region $r>\overline{r}$:\\
a) during the first few seconds (at least for stars initially cold and having masses $\lesssim 1.5 M_{\odot}$) the neutrino luminosity is dominated by the cooling of the inner region  $r<\overline{r}$\\
b) neutrino luminosity dominated by the cooling of the outer region $r>\overline{r}$. 
This phase lasts a few tens of seconds and displays a quasi-plateau which originates from 
the burning velocity which is inversely proportional to the temperature.
\end{itemize}

The quasi-plateau produced during the phase b) has a rather large
luminosity $\sim 10^{51-52}$ erg/s for tens of seconds and, if
detected, would be a unique feature of the conversion of an hadronic
star into a quark star. This very interesting possibility deserves a
more refined treatment of the cooling process possibly via a neutrino
transport code.


The prolonged conversion of the star is very promising also in
connection with the mechanism generating long gamma-ray-bursts within
the protomagnetar model \cite{Metzger:2010pp}. This model needs three
crucial ingredients: high rotation frequency, high magnetic field and
a significant neutrino emission (provided by the cooling of the
protomagnetar) which, through ablation of the external layers of the
compact star, allows to obtain the correct value of the Lorentz factor
of the wind within which the gamma-ray-burst is generated.  The
neutrino emission caused by the conversion of an old hadronic star
could lead to a long gamma-ray-burst if the other two requirements
(i.e. high magnetic filed and high rotation frequency) are
fulfilled. One possibility is given by a merger of a neutron star with
a white dwarf. It has been shown that this process produces a spinning
Thorne-Zytkow-like object with a low temperature, $T\sim 10^9$ K
\cite{Paschalidis:2011ez}.  If large magnetic fields are generated,
for instance via magnetorotational instabilities, the conditions for
producing a gamma-ray-burst are fulfilled.  The accretion of matter
onto the hadronic star would trigger the conversion to a quark star
and the expected neutrino luminosities are similar to the ones
presented in Fig. 4.  Such a gamma-ray-burst would be similar to a
short gamma-ray-burst because it is associated with the merger of two
compact stars but its duration would be comparable to the one of long
gamma-ray-bursts.  These features are in agreement with the analyses
of GRB060614 \cite{Fynbo:2006mw,DellaValle:2006nf,GalYam:2006en} where
it has been argued that this burst is not associated with a supernova.

\vskip 1cm

G.P. acknowledges financial support from the Italian Ministry of
Research through the program \textquotedblleft Rita Levi
Montalcini\textquotedblright. The authors gratefully acknowledge the COST Action
MP1304 "NewCompStar" for supporting their networking and collaboration activities.

\newpage

\section{Appendix A}
We generalize the discussion presented in Sec.II, where we have adopted
the quark matter equation of state with massless quarks,
to the case in which the equation of state of fluid 2 is a generic 
polytrope given by:
$e_q=\alpha n_q + p_q/(\gamma-1)$ (\cite{Ozel:2010fw}), $p_q=k n_q^\gamma$, where $\gamma$ is the adiabatic index
and $1<\gamma \leq 2$ (the second inequality implying that the equation of state is causal at all densities \cite{Read:2008iy}). 

One can derive the following expression for the energy 
density as a function of $p_q$ and $X_q$:
\begin{equation}
e_q(p_q,X_q)=\frac{\alpha^2(\gamma-1)+2p_qX_q+\alpha\sqrt{\gamma-1}\sqrt{\alpha^2(\gamma-1)+4\gamma p_qX_q}}{2X_q(\gamma-1)}
\label{formulapp}
\end{equation}

Let us fix $X_B=X_A$ and assume $\Delta(p_A,X_A)>0$.
If $p_B > p_A$, the initial point A lies in the region of the (p,X) plane below the detonation adiabat.
The detonation adiabat reads (adding and subtracting $e_q(p_A,X_A)$):
\begin{equation}
\Delta(p_A,X_A)=p_A-p_B+e_q(p_B,X_A)-e_q(p_A,X_A)
\end{equation}
which after some manipulation and using Eq.\ref{formulapp} reads:
\begin{eqnarray}
\Delta(p_A,X_A)&=&\frac{\alpha}{2X_A\sqrt{\gamma-1}}\left(\sqrt{\alpha^2(\gamma-1)+4\gamma p_B X_A}
-\sqrt{\alpha^2(\gamma-1)+4\gamma p_A X_A}\right)\\&+&(p_B-p_A)\frac{2-\gamma}{\gamma-1}
\end{eqnarray}
Since $1<\gamma \leq 2$ the sign of $\Delta(p_A,X_A)$ clearly determines the sign of $p_B-p_A$.
Thus, if $\Delta(p_A,X_A)>0$, i.e. if the Coll's condition holds true,
the initial point A lies in the region of the (p,X) plane below the detonation adiabat.
One sees again the Coll's condition establishes the position of the initial state
of the hadronic phase with respect to the detonation adiabat.
\footnote{For values of $\gamma>2$, one can still obtain the same 
conclusion but depending on the specific values of $\alpha$, $p_A$ and $X_A$.}

\section{Appendix B}
In this section we aim at providing an estimate on the allowed values of 
the parameter $a_{Q*}\equiv (n_d(0^+)-n_s(0^+))/(2 n_Q)$. Please notice that
in this definition, appear quantities computed at the interface and a quantity
computed for $x\rightarrow +\infty$.
In particular we estimate $a_{Q*}^{max}$
which is the maximum value of the strangeness unbalance 
at the interface between hadronic matter and quark matter
in order for the process of conversion to still proceed because of its exothermicity, see Fig.4 of Ref.~\cite{Alford:2014jha}. 
We assume, as in  Ref.~\cite{Alford:2014jha}, that the deconfinement always takes place 
with maximal velocity i.e. with  $a_{Q*}=a_{Q*}^{max}$.
We consider the case of a bag-model 
equation of state with massless quarks and we limit our discussion to the case of vanishing pressure $p$.
This case is particular relevant for the burning of the most external layers of the hadronic star
where the pressure approaches zero.
Following the scheme of Ref.~\cite{Alberico:2001zz},
we introduce the parameter $r_s$ which is the ratio between the baryonic density of strange quarks
and the baryon density: $r_s=n_s/3n$. The Fermi momenta of 
up, down and strange quarks read: $k_{F_{u,d}}(n,r_s)=k_F(n,r_s)=(1.5\, \pi^2 n (1-r_s))^{1/3}$,
$k_{F_s}(n,r_s)=(3\pi^2 (r_s n))^{1/3}$ where we have assumed that 
up and down quarks have the same Fermi momenta (as in beta stable quark matter 
for the case of massless quarks). The energy density reads:
\begin{equation}
e=\frac{6 k_F^4}{4\pi^2}+\frac{3 k_{F_s}^4}{4\pi^2}+B.
\end{equation} 
The bag constant $B$ varies in the range $B_{min}<B<B_{max}$ where for $B=B_{max}$
the energy per baryon of strange quark matter is $930$ MeV and 
for $B=B_{min}$ the energy per baryon of two flavor quark matter is $930$ MeV. 

We first compute $n_Q$ which corresponds 
to finding the baryon density $n=n_Q$ for which beta stable quark matter (i.e. $r_s=1/3$) 
has vanishing pressure. At this value of density the energy per baryon 
$(e/n)_{\beta}$ is strictly smaller than $930$ MeV because we are assuming
the hypothesis of 
absolutely stable strange quark matter.

We now calculate the maximum value of $n_d(0^+)-n_s(0^+)$.  We define
$\overline{n}$ and $\overline{r_s}$ as the values of $n$ and $r_s$ for
which the pressure $p=0$ and the energy per baryon of non beta-stable quark matter $(e/n)_{non-\beta}=930$ MeV.
$\overline{r_s}$ represents therefore the minimum amount of
strangeness for which non-beta stable strange quark matter has an
energy per baryon equal to the one of iron.
A quark phase with $r_s > \overline{r_s}$  would be favored with respect
to nuclear matter and the conversion process would therefore be exothermic.

$a_{Q*}^{max}$ is related to $\overline{r_s}$ and $\overline{n}$ 
by the following equation:
\begin{equation}
a_{Q*}^{max}=\frac{k_F^3(\overline{n},\overline{r_s})-
k_{F_s}^3(\overline{n},\overline{r_s})}{2 \pi^2 n_Q}
\label{eq-last}
\end{equation}
For each value of $B$, $\overline{r_s}$, $\overline{n}$
and $n_Q$ are completely determined and thus 
also $a_{Q*}^{max}$ is fixed. 

\begin{figure}[htb]
\vskip 2cm
\begin{centering}
\epsfig{file=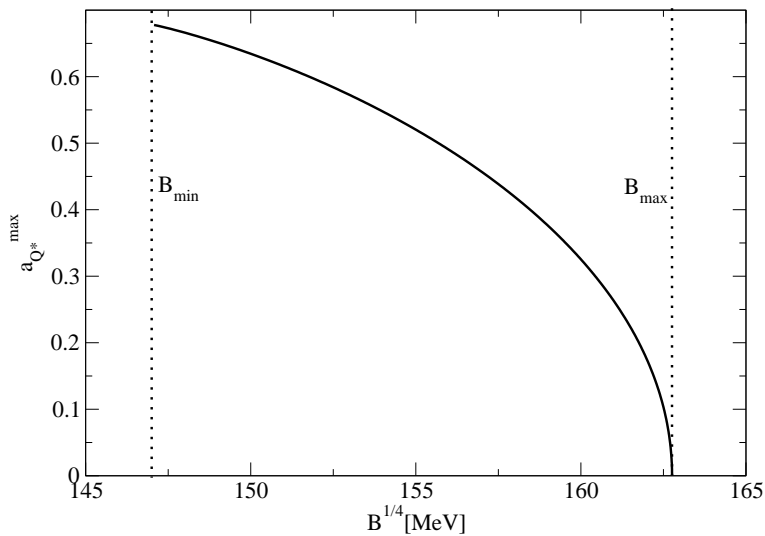,height=10cm,width=7cm,angle=-90}
\caption{Dependence of $a_{Q*}^{max}$ on the bag constant.}
\label{appendice}
\end{centering}
\end{figure}

We display the results in Fig. \ref{appendice}. 
In general  $a_{Q*}^{max} <1$, its maximum value equals 1 only if one considers for $x\rightarrow +\infty$ non beta-stable 2 flavor quark matter
with $n_d=2n_u$ (see \cite{1987PhLB..192...71O}). We consider instead, as in \cite{Alford:2014jha},  beta stable strange quark matter for $x\rightarrow +\infty$. 
One can show analytically that in this case the maximum value of $a_{Q*}^{max}$ is  $1.5^{3/4}/2 \sim 0.667$ 
and it is obtained for $B=B_{min}$ \footnote{This value is obtained by using 
the formulae for the pressure of two flavor and three flavor quark matter: $p_{2F}=k_F^4/(2 \pi^2)-B$ and  $p_{3F}=3k_F^4/(4 \pi^2)-B$ where $k_F$ is the quark Fermi momentum which is the same for all flavors. By imposing  $p_{2F}=p_{3F}=0$ one obtains the quark Fermi momentum for $x=0$ and for $x\rightarrow +\infty$. Finally one uses \ref{eq-last} to obtain 
$a_{Q*}^{max}$. }.

The minimum value of $a_{Q*}^{max}$ is zero and it is
obtained at $B=B_{max}$. Notice that for this value of $B$, $\frac{\mathrm{d}a_{Q*}^{max}}{\mathrm{d}B} \rightarrow \infty$.
This basically implies that the parameter space (i.e. the range of $B$) has little room for the minimum value of $a_{Q*}^{max}$.
In other words, it is very unlikely that  $a_{Q*}^{max}\sim 0$. In turns, this means that while the value of the velocity of conversion
is quite uncertain it cannot be vanishingly small.

\begin{thebibliography}{38}
\expandafter\ifx\csname natexlab\endcsname\relax\def\natexlab#1{#1}\fi
\expandafter\ifx\csname bibnamefont\endcsname\relax
  \def\bibnamefont#1{#1}\fi
\expandafter\ifx\csname bibfnamefont\endcsname\relax
  \def\bibfnamefont#1{#1}\fi
\expandafter\ifx\csname citenamefont\endcsname\relax
  \def\citenamefont#1{#1}\fi
\expandafter\ifx\csname url\endcsname\relax
  \def\url#1{\texttt{#1}}\fi
\expandafter\ifx\csname urlprefix\endcsname\relax\def\urlprefix{URL }\fi
\providecommand{\bibinfo}[2]{#2}
\providecommand{\eprint}[2][]{\url{#2}}

\bibitem[{\citenamefont{Bodmer}(1971)}]{Bodmer:1971we}
\bibinfo{author}{\bibfnamefont{A.}~\bibnamefont{Bodmer}},
  \bibinfo{journal}{Phys.Rev.} \textbf{\bibinfo{volume}{D4}},
  \bibinfo{pages}{1601} (\bibinfo{year}{1971}).

\bibitem[{\citenamefont{{Witten}}(1984)}]{1984PhRvD..30..272W}
\bibinfo{author}{\bibfnamefont{E.}~\bibnamefont{{Witten}}},
  \bibinfo{journal}{\prd} \textbf{\bibinfo{volume}{30}}, \bibinfo{pages}{272}
  (\bibinfo{year}{1984}).

\bibitem[{\citenamefont{Alcock et~al.}(1986)\citenamefont{Alcock, Farhi, and
  Olinto}}]{Alcock:1986hz}
\bibinfo{author}{\bibfnamefont{C.}~\bibnamefont{Alcock}},
  \bibinfo{author}{\bibfnamefont{E.}~\bibnamefont{Farhi}}, \bibnamefont{and}
  \bibinfo{author}{\bibfnamefont{A.}~\bibnamefont{Olinto}},
  \bibinfo{journal}{Astrophys.J.} \textbf{\bibinfo{volume}{310}},
  \bibinfo{pages}{261} (\bibinfo{year}{1986}).

\bibitem[{\citenamefont{Haensel et~al.}(1986)\citenamefont{Haensel, Zdunik, and
  Schaeffer}}]{Haensel:1986qb}
\bibinfo{author}{\bibfnamefont{P.}~\bibnamefont{Haensel}},
  \bibinfo{author}{\bibfnamefont{J.}~\bibnamefont{Zdunik}}, \bibnamefont{and}
  \bibinfo{author}{\bibfnamefont{R.}~\bibnamefont{Schaeffer}},
  \bibinfo{journal}{Astron.Astrophys.} \textbf{\bibinfo{volume}{160}},
  \bibinfo{pages}{121} (\bibinfo{year}{1986}).

\bibitem[{\citenamefont{Madsen}(1988)}]{Madsen:1989pg}
\bibinfo{author}{\bibfnamefont{J.}~\bibnamefont{Madsen}},
  \bibinfo{journal}{Phys.Rev.Lett.} \textbf{\bibinfo{volume}{61}},
  \bibinfo{pages}{2909} (\bibinfo{year}{1988}).

\bibitem[{\citenamefont{{Olinto}}(1987)}]{1987PhLB..192...71O}
\bibinfo{author}{\bibfnamefont{A.~V.} \bibnamefont{{Olinto}}},
  \bibinfo{journal}{Physics Letters B} \textbf{\bibinfo{volume}{192}},
  \bibinfo{pages}{71} (\bibinfo{year}{1987}).

\bibitem[{\citenamefont{Williams}(1985)}]{williams}
\bibinfo{author}{\bibfnamefont{F.}~\bibnamefont{Williams}},
  \emph{\bibinfo{title}{{Combustion theory}}} (\bibinfo{publisher}{The
  Benjamin/Cummings Publishing Company, Inc.}, \bibinfo{address}{Menlo Park,
  California (USA)}, \bibinfo{year}{1985}).

\bibitem[{\citenamefont{{Niebergal} et~al.}(2010)\citenamefont{{Niebergal},
  {Ouyed}, and {Jaikumar}}}]{niebergal2010a}
\bibinfo{author}{\bibfnamefont{B.}~\bibnamefont{{Niebergal}}},
  \bibinfo{author}{\bibfnamefont{R.}~\bibnamefont{{Ouyed}}}, \bibnamefont{and}
  \bibinfo{author}{\bibfnamefont{P.}~\bibnamefont{{Jaikumar}}},
  \bibinfo{journal}{\prc} \textbf{\bibinfo{volume}{82}},
  \bibinfo{pages}{062801} (\bibinfo{year}{2010}).

\bibitem[{\citenamefont{Landau and Lifshitz}(1987)}]{landau}
\bibinfo{author}{\bibfnamefont{L.~D.} \bibnamefont{Landau}} \bibnamefont{and}
  \bibinfo{author}{\bibfnamefont{E.~M.} \bibnamefont{Lifshitz}},
  \emph{\bibinfo{title}{{Fluid Mechanics, Second Edition: Volume 6 (Course of
  Theoretical Physics)}}}, Course of theoretical physics / by L. D. Landau and
  E. M. Lifshitz, Vol. 6 (\bibinfo{publisher}{Butterworth-Heinemann},
  \bibinfo{year}{1987}), \bibinfo{edition}{2nd} ed., ISBN
  \bibinfo{isbn}{0750627670}.

\bibitem[{\citenamefont{Alford et~al.}(2015)\citenamefont{Alford, Han, and
  Schwenzer}}]{Alford:2014jha}
\bibinfo{author}{\bibfnamefont{M.~G.} \bibnamefont{Alford}},
  \bibinfo{author}{\bibfnamefont{S.}~\bibnamefont{Han}}, \bibnamefont{and}
  \bibinfo{author}{\bibfnamefont{K.}~\bibnamefont{Schwenzer}},
  \bibinfo{journal}{Phys.Rev.} \textbf{\bibinfo{volume}{C91}},
  \bibinfo{pages}{055804} (\bibinfo{year}{2015}), \eprint{1404.5279}.

\bibitem[{\citenamefont{Reinecke et~al.}(1999)\citenamefont{Reinecke,
  Hillebrandt, Niemeyer, Klein, and Groebl}}]{Reinecke:1998mk}
\bibinfo{author}{\bibfnamefont{M.}~\bibnamefont{Reinecke}},
  \bibinfo{author}{\bibfnamefont{W.}~\bibnamefont{Hillebrandt}},
  \bibinfo{author}{\bibfnamefont{J.~C.} \bibnamefont{Niemeyer}},
  \bibinfo{author}{\bibfnamefont{R.}~\bibnamefont{Klein}}, \bibnamefont{and}
  \bibinfo{author}{\bibfnamefont{A.}~\bibnamefont{Groebl}},
  \bibinfo{journal}{Astron.Astrophys.} \textbf{\bibinfo{volume}{347}},
  \bibinfo{pages}{724} (\bibinfo{year}{1999}), \eprint{astro-ph/9812119}.

\bibitem[{\citenamefont{Hillebrandt and Niemeyer}(2000)}]{Hillebrandt:2000ga}
\bibinfo{author}{\bibfnamefont{W.}~\bibnamefont{Hillebrandt}} \bibnamefont{and}
  \bibinfo{author}{\bibfnamefont{J.~C.} \bibnamefont{Niemeyer}},
  \bibinfo{journal}{Ann.Rev.Astron.Astrophys.} \textbf{\bibinfo{volume}{38}},
  \bibinfo{pages}{191} (\bibinfo{year}{2000}), \eprint{astro-ph/0006305}.

\bibitem[{\citenamefont{Blinnikov et~al.}(1995)\citenamefont{Blinnikov,
  Sasorov, and Woosley}}]{blinnikov}
\bibinfo{author}{\bibfnamefont{S.~I.} \bibnamefont{Blinnikov}},
  \bibinfo{author}{\bibfnamefont{P.~V.} \bibnamefont{Sasorov}},
  \bibnamefont{and} \bibinfo{author}{\bibfnamefont{S.~E.}
  \bibnamefont{Woosley}}, \bibinfo{journal}{Space Science Reviews}
  \textbf{\bibinfo{volume}{74}}, \bibinfo{pages}{299} (\bibinfo{year}{1995}).

\bibitem[{\citenamefont{Lugones et~al.}(1994)\citenamefont{Lugones, Benvenuto,
  and Vucetich}}]{Lugones:1994xg}
\bibinfo{author}{\bibfnamefont{G.}~\bibnamefont{Lugones}},
  \bibinfo{author}{\bibfnamefont{O.}~\bibnamefont{Benvenuto}},
  \bibnamefont{and} \bibinfo{author}{\bibfnamefont{H.}~\bibnamefont{Vucetich}},
  \bibinfo{journal}{Phys.Rev.} \textbf{\bibinfo{volume}{D50}},
  \bibinfo{pages}{6100} (\bibinfo{year}{1994}).

\bibitem[{\citenamefont{Horvath}(2010{\natexlab{a}})}]{Horvath:2007tv}
\bibinfo{author}{\bibfnamefont{J.}~\bibnamefont{Horvath}},
  \bibinfo{journal}{Int.J.Mod.Phys.} \textbf{\bibinfo{volume}{D19}},
  \bibinfo{pages}{523} (\bibinfo{year}{2010}{\natexlab{a}}),
  \eprint{astro-ph/0703233}.

\bibitem[{\citenamefont{Horvath}(2010{\natexlab{b}})}]{Horvath:2010sc}
\bibinfo{author}{\bibfnamefont{J.}~\bibnamefont{Horvath}}
  (\bibinfo{year}{2010}{\natexlab{b}}), \eprint{1005.4302}.

\bibitem[{\citenamefont{Drago et~al.}(2007)\citenamefont{Drago, Lavagno, and
  Parenti}}]{Drago:2005yj}
\bibinfo{author}{\bibfnamefont{A.}~\bibnamefont{Drago}},
  \bibinfo{author}{\bibfnamefont{A.}~\bibnamefont{Lavagno}}, \bibnamefont{and}
  \bibinfo{author}{\bibfnamefont{I.}~\bibnamefont{Parenti}},
  \bibinfo{journal}{Astrophys.J.} \textbf{\bibinfo{volume}{659}},
  \bibinfo{pages}{1519} (\bibinfo{year}{2007}).

\bibitem[{\citenamefont{Herzog and R{\"o}pke}(2011)}]{Herzog:2011sn}
\bibinfo{author}{\bibfnamefont{M.}~\bibnamefont{Herzog}} \bibnamefont{and}
  \bibinfo{author}{\bibfnamefont{F.~K.} \bibnamefont{R{\"o}pke}},
  \bibinfo{journal}{Phys.Rev.} \textbf{\bibinfo{volume}{D84}},
  \bibinfo{pages}{083002} (\bibinfo{year}{2011}).

\bibitem[{\citenamefont{Pagliara et~al.}(2013)\citenamefont{Pagliara, Herzog,
  and Ropke}}]{Pagliara:2013tza}
\bibinfo{author}{\bibfnamefont{G.}~\bibnamefont{Pagliara}},
  \bibinfo{author}{\bibfnamefont{M.}~\bibnamefont{Herzog}}, \bibnamefont{and}
  \bibinfo{author}{\bibfnamefont{F.}~\bibnamefont{Ropke}},
  \bibinfo{journal}{Phys.Rev.,} \textbf{\bibinfo{volume}{D87}},
  \bibinfo{pages}{103007} (\bibinfo{year}{2013}).

\bibitem[{\citenamefont{Coll}(1976)}]{coll}
\bibinfo{author}{\bibfnamefont{B.}~\bibnamefont{Coll}},
  \bibinfo{journal}{Annales de l'I.H.P. A} \textbf{\bibinfo{volume}{25}},
  \bibinfo{pages}{363} (\bibinfo{year}{1976}).

\bibitem[{\citenamefont{Anile}(1990)}]{anile}
\bibinfo{author}{\bibfnamefont{A.}~\bibnamefont{Anile}},
  \emph{\bibinfo{title}{Relativistic Fluids and Magneto-fluids}}
  (\bibinfo{publisher}{Cambridge University Press}, \bibinfo{year}{1990}).

\bibitem[{\citenamefont{Zeldovich}(1992)}]{zeldo}
\bibinfo{author}{\bibfnamefont{Y.}~\bibnamefont{Zeldovich}},
  \emph{\bibinfo{title}{{Selected works of Y.B. Zeldovich}}}
  (\bibinfo{publisher}{Princeton University Press},
  \bibinfo{address}{Princeton, New Jersey (USA)}, \bibinfo{year}{1992}).

\bibitem[{\citenamefont{Drago and Pagliara}(2015)}]{Drago:2015gsa}
\bibinfo{author}{\bibfnamefont{A.}~\bibnamefont{Drago}} \bibnamefont{and}
  \bibinfo{author}{\bibfnamefont{G.}~\bibnamefont{Pagliara}}
  (\bibinfo{year}{2015}), \eprint{1504.02795}.

\bibitem[{\citenamefont{Lattimer and Swesty}(1991)}]{Lattimer:1991nc}
\bibinfo{author}{\bibfnamefont{J.~M.} \bibnamefont{Lattimer}} \bibnamefont{and}
  \bibinfo{author}{\bibfnamefont{F.~D.} \bibnamefont{Swesty}},
  \bibinfo{journal}{Nucl.Phys.} \textbf{\bibinfo{volume}{A535}},
  \bibinfo{pages}{331} (\bibinfo{year}{1991}).

\bibitem[{\citenamefont{Glendenning and Moszkowski}(1991)}]{Glendenning:1991es}
\bibinfo{author}{\bibfnamefont{N.}~\bibnamefont{Glendenning}} \bibnamefont{and}
  \bibinfo{author}{\bibfnamefont{S.}~\bibnamefont{Moszkowski}},
  \bibinfo{journal}{Phys.Rev.Lett.} \textbf{\bibinfo{volume}{67}},
  \bibinfo{pages}{2414} (\bibinfo{year}{1991}).

\bibitem[{\citenamefont{Steiner et~al.}(2013)\citenamefont{Steiner, Hempel, and
  Fischer}}]{Steiner:2012rk}
\bibinfo{author}{\bibfnamefont{A.~W.} \bibnamefont{Steiner}},
  \bibinfo{author}{\bibfnamefont{M.}~\bibnamefont{Hempel}}, \bibnamefont{and}
  \bibinfo{author}{\bibfnamefont{T.}~\bibnamefont{Fischer}},
  \bibinfo{journal}{Astrophys.J.} \textbf{\bibinfo{volume}{774}},
  \bibinfo{pages}{17} (\bibinfo{year}{2013}), \eprint{1207.2184}.

\bibitem[{\citenamefont{Drago et~al.}(2014{\natexlab{a}})\citenamefont{Drago,
  Lavagno, Pagliara, and Pigato}}]{Drago:2014oja}
\bibinfo{author}{\bibfnamefont{A.}~\bibnamefont{Drago}},
  \bibinfo{author}{\bibfnamefont{A.}~\bibnamefont{Lavagno}},
  \bibinfo{author}{\bibfnamefont{G.}~\bibnamefont{Pagliara}}, \bibnamefont{and}
  \bibinfo{author}{\bibfnamefont{D.}~\bibnamefont{Pigato}}
  (\bibinfo{year}{2014}{\natexlab{a}}), \eprint{1407.2843}.

\bibitem[{\citenamefont{Drago et~al.}(2014{\natexlab{b}})\citenamefont{Drago,
  Lavagno, and Pagliara}}]{Drago:2013fsa}
\bibinfo{author}{\bibfnamefont{A.}~\bibnamefont{Drago}},
  \bibinfo{author}{\bibfnamefont{A.}~\bibnamefont{Lavagno}}, \bibnamefont{and}
  \bibinfo{author}{\bibfnamefont{G.}~\bibnamefont{Pagliara}},
  \bibinfo{journal}{Phys.Rev.} \textbf{\bibinfo{volume}{D89}},
  \bibinfo{pages}{043014} (\bibinfo{year}{2014}{\natexlab{b}}),
  \eprint{1309.7263}.

\bibitem[{\citenamefont{Haensel and Zdunik}(2006)}]{Haensel:2007tf}
\bibinfo{author}{\bibfnamefont{P.}~\bibnamefont{Haensel}} \bibnamefont{and}
  \bibinfo{author}{\bibfnamefont{J.}~\bibnamefont{Zdunik}},
  \bibinfo{journal}{Nuovo Cim.} \textbf{\bibinfo{volume}{B121}},
  \bibinfo{pages}{1349} (\bibinfo{year}{2006}), \eprint{astro-ph/0701258}.

\bibitem[{\citenamefont{Shapiro and Teukolsky}(1983)}]{shapiro}
\bibinfo{author}{\bibfnamefont{S.}~\bibnamefont{Shapiro}} \bibnamefont{and}
  \bibinfo{author}{\bibfnamefont{S.}~\bibnamefont{Teukolsky}},
  \emph{\bibinfo{title}{{Black holes, white dwarfs, and neutron stars: The
  physics of compact objects}}} (\bibinfo{publisher}{Wiley \& Sons},
  \bibinfo{address}{New York, Chichester Brisbane, Toronto, Singapore},
  \bibinfo{year}{1983}).

\bibitem[{\citenamefont{Metzger et~al.}(2010)\citenamefont{Metzger, Giannios,
  Thompson, Bucciantini, and Quataert}}]{Metzger:2010pp}
\bibinfo{author}{\bibfnamefont{B.}~\bibnamefont{Metzger}},
  \bibinfo{author}{\bibfnamefont{D.}~\bibnamefont{Giannios}},
  \bibinfo{author}{\bibfnamefont{T.}~\bibnamefont{Thompson}},
  \bibinfo{author}{\bibfnamefont{N.}~\bibnamefont{Bucciantini}},
  \bibnamefont{and} \bibinfo{author}{\bibfnamefont{E.}~\bibnamefont{Quataert}}
  (\bibinfo{year}{2010}).

\bibitem[{\citenamefont{Paschalidis et~al.}(2011)\citenamefont{Paschalidis,
  Liu, Etienne, and Shapiro}}]{Paschalidis:2011ez}
\bibinfo{author}{\bibfnamefont{V.}~\bibnamefont{Paschalidis}},
  \bibinfo{author}{\bibfnamefont{Y.~T.} \bibnamefont{Liu}},
  \bibinfo{author}{\bibfnamefont{Z.}~\bibnamefont{Etienne}}, \bibnamefont{and}
  \bibinfo{author}{\bibfnamefont{S.~L.} \bibnamefont{Shapiro}},
  \bibinfo{journal}{Phys.Rev.} \textbf{\bibinfo{volume}{D84}},
  \bibinfo{pages}{104032} (\bibinfo{year}{2011}), \eprint{1109.5177}.

\bibitem[{\citenamefont{Fynbo et~al.}(2006)\citenamefont{Fynbo, Watson, Thoene,
  Sollerman, Bloom et~al.}}]{Fynbo:2006mw}
\bibinfo{author}{\bibfnamefont{J.~P.} \bibnamefont{Fynbo}},
  \bibinfo{author}{\bibfnamefont{D.}~\bibnamefont{Watson}},
  \bibinfo{author}{\bibfnamefont{C.~C.} \bibnamefont{Thoene}},
  \bibinfo{author}{\bibfnamefont{J.}~\bibnamefont{Sollerman}},
  \bibinfo{author}{\bibfnamefont{J.~S.} \bibnamefont{Bloom}},
  \bibnamefont{et~al.}, \bibinfo{journal}{Nature}
  \textbf{\bibinfo{volume}{444}}, \bibinfo{pages}{1047} (\bibinfo{year}{2006}),
  \eprint{astro-ph/0608313}.

\bibitem[{\citenamefont{Della~Valle et~al.}(2006)\citenamefont{Della~Valle,
  Chincarini, Panagia, Tagliaferri, Malesani et~al.}}]{DellaValle:2006nf}
\bibinfo{author}{\bibfnamefont{M.}~\bibnamefont{Della~Valle}},
  \bibinfo{author}{\bibfnamefont{G.}~\bibnamefont{Chincarini}},
  \bibinfo{author}{\bibfnamefont{N.}~\bibnamefont{Panagia}},
  \bibinfo{author}{\bibfnamefont{G.}~\bibnamefont{Tagliaferri}},
  \bibinfo{author}{\bibfnamefont{D.}~\bibnamefont{Malesani}},
  \bibnamefont{et~al.}, \bibinfo{journal}{Nature}
  \textbf{\bibinfo{volume}{444}}, \bibinfo{pages}{1050} (\bibinfo{year}{2006}),
  \eprint{astro-ph/0608322}.

\bibitem[{\citenamefont{Gal-Yam et~al.}(2006)\citenamefont{Gal-Yam, Fox, Price,
  Davis, Leonard et~al.}}]{GalYam:2006en}
\bibinfo{author}{\bibfnamefont{A.}~\bibnamefont{Gal-Yam}},
  \bibinfo{author}{\bibfnamefont{D.}~\bibnamefont{Fox}},
  \bibinfo{author}{\bibfnamefont{P.}~\bibnamefont{Price}},
  \bibinfo{author}{\bibfnamefont{M.}~\bibnamefont{Davis}},
  \bibinfo{author}{\bibfnamefont{D.}~\bibnamefont{Leonard}},
  \bibnamefont{et~al.}, \bibinfo{journal}{Nature}
  \textbf{\bibinfo{volume}{444}}, \bibinfo{pages}{1053} (\bibinfo{year}{2006}),
  \eprint{astro-ph/0608257}.

\bibitem[{\citenamefont{Ozel et~al.}(2010)\citenamefont{Ozel, Baym, and
  Guver}}]{Ozel:2010fw}
\bibinfo{author}{\bibfnamefont{F.}~\bibnamefont{Ozel}},
  \bibinfo{author}{\bibfnamefont{G.}~\bibnamefont{Baym}}, \bibnamefont{and}
  \bibinfo{author}{\bibfnamefont{T.}~\bibnamefont{Guver}},
  \bibinfo{journal}{Phys.Rev.} \textbf{\bibinfo{volume}{D82}},
  \bibinfo{pages}{101301} (\bibinfo{year}{2010}), \eprint{1002.3153}.

\bibitem[{\citenamefont{Read et~al.}(2009)\citenamefont{Read, Lackey, Owen, and
  Friedman}}]{Read:2008iy}
\bibinfo{author}{\bibfnamefont{J.~S.} \bibnamefont{Read}},
  \bibinfo{author}{\bibfnamefont{B.~D.} \bibnamefont{Lackey}},
  \bibinfo{author}{\bibfnamefont{B.~J.} \bibnamefont{Owen}}, \bibnamefont{and}
  \bibinfo{author}{\bibfnamefont{J.~L.} \bibnamefont{Friedman}},
  \bibinfo{journal}{Phys.Rev.} \textbf{\bibinfo{volume}{D79}},
  \bibinfo{pages}{124032} (\bibinfo{year}{2009}), \eprint{0812.2163}.

\bibitem[{\citenamefont{Alberico et~al.}(2002)\citenamefont{Alberico, Drago,
  and Ratti}}]{Alberico:2001zz}
\bibinfo{author}{\bibfnamefont{W.}~\bibnamefont{Alberico}},
  \bibinfo{author}{\bibfnamefont{A.}~\bibnamefont{Drago}}, \bibnamefont{and}
  \bibinfo{author}{\bibfnamefont{C.}~\bibnamefont{Ratti}},
  \bibinfo{journal}{Nucl.Phys.} \textbf{\bibinfo{volume}{A706}},
  \bibinfo{pages}{143} (\bibinfo{year}{2002}), \eprint{hep-ph/0110091}.

\end{thebibliography}

\end{document}